\documentclass[12pt, journal, onecolumn]{IEEEtran}

\usepackage{epsfig,cite}
\usepackage{setspace}
\usepackage{subfigure}

\usepackage{graphicx}
\usepackage{longtable}
\usepackage{amsmath}
\usepackage{color}
\usepackage[finnish,english]{babel}
\usepackage[T1]{fontenc}

\def\e{\begin{equation}}
\def\f{\end{equation}}
\def\%#1{\mbox{\boldmath $#1$}}
\def\=#1{\overline{\overline #1}}

\def\_#1{{\bf #1}}

\def\.{\cdot}

\def\##1{{\bf#1\mit}}

\def\l#1{\label{eq:#1}}
\def\r#1{(\ref{eq:#1})}
\def\am{\left(\begin{array}{c}}
\def\amm{\left(\begin{array}{cc}}
\def\a{\end{array}\right)}

\title{Simple and Accurate Analytical Model of Planar Grids and High-Impedance Surfaces Comprising Metal Strips or Patches}

\author{Olli Luukkonen, Constantin Simovski, \IEEEmembership{Member,~IEEE}, G\'erard Granet,\\ George Goussetis,~\IEEEmembership{Member,~IEEE}, Dmitri Lioubtchenko, Antti
V.~R\"ais\"anen,~\IEEEmembership{Fellow,~IEEE},\\ and Sergei
A.~Tretyakov,~\IEEEmembership{Fellow,~IEEE}
\thanks{The work was supported in part by the Academy of
Finland and Tekes through the Center-of-Excellence Programme.}
\thanks{O.~Luukkonen, C.~Simovski, D.~Lioubtchenko, A.~V.~R\"ais\"anen
and S.~A.~Tretyakov are with Department of Radio Science and
Engineering/SMARAD CoE, TKK Helsinki University of Technology, P.O.
3000, FI-02015 TKK, Finland (email: olli.luukkonen@tkk.fi)}
\thanks{C.~Simovski is also with Department of Physics,
St.~Petersburg Institute of Fine Mechanics and Optics, 197101,
Sablinskaya 14, St.~Petersburg, Russia.}
\thanks{G.~Granet is with LASMEA, Universit\'e Blaise Pascal / CNRS,
F-63177 Aubi\'ere Cedex, France} \thanks{G.~Goussetis is with the
Institute of Integrated Systems, Edinburgh Research Partnership,
Heriot-Watt University Edinburgh, EH14 AS, U.K.}}

\begin{document}

\maketitle {\center \large }

\parskip 0pt

\begin{abstract}

This paper introduces simple analytical formulas for the grid
impedance of electrically dense arrays of square patches and for the
surface impedance of high-impedance surfaces based on the dense
arrays of metal strips or square patches over ground planes.
Emphasis is on the oblique-incidence excitation. The approach is
based on the known analytical models for strip grids combined with
the approximate Babinet principle for planar grids located at a
dielectric interface. Analytical expressions for the surface
impedance and reflection coefficient resulting from our analysis are
thoroughly verified by full-wave simulations and compared with
available data in open literature for particular cases. The results
can be used in the design of various antennas and microwave or
millimeter wave devices which use artificial impedance surfaces and
artificial magnetic conductors (reflect-array antennas, tunable
phase shifters, etc.), as well as for the derivation of accurate
higher-order impedance boundary conditions for artificial (high-)
impedance surfaces. As an example, the propagation properties of
surface waves along the high-impedance surfaces are studied.

\end{abstract}

\section{Introduction}

In this paper we consider planar periodic arrays of infinitely long
metal strips and periodic arrays of square patches, as well as
artificial high-impedance surfaces based on such grids. Possible
applications for these arrays include artificial dielectrics
\cite{Cohn}, antenna radomes \cite{Chen_transmission}, other
applications typical for frequency selective surfaces \cite{Munk},
and artificial high-impedance surfaces
\cite{Sievenpiper,Tretyakov_1,Simovski_1,Simovski, Maci2}, where
such arrays are located on a metal-backed dielectric layer (which
may be perforated with metal vias). Furthermore, possible
applications for artificial impedance surfaces expand the list to
phase shifters \cite{Chicherin,Higgins}, TEM waveguides \cite{Yang},
planar reflect-arrays \cite{Sievenpiper_beamsteering}, absorbers
\cite{engheta,Tretyakov_motl,gao,simms}, and artificial magnetic
conductors (engineered antenna ground planes)
\cite{Sievenpiper_thesis}. Capacitive strips and square patches have
been studied extensively in the literature (e.g.,
\cite{Marcuvitz,Lee,Goussetis}). However, to the best of the
authors' knowledge, there is no known easily applicable analytical
model capable of predicting the plane-wave response of these
artificial surfaces for large angles of incidence with good
accuracy.

Models of planar arrays of metal elements excited by plane waves can
be roughly split into two categories: computational and analytical
methods. Computational methods as a rule are based on the Floquet
expansion of the scattered field (see, e.g.,
\cite{Chen_transmission,Munk,Maci,Whites}). These methods are
electromagnetically strict and general (i.e., not restricted to a
particular design geometry). Periodicity of the total field in
tangential directions allows one to consider the incidence of a
plane wave on a planar grid or on a high-impedance surface as a
single unit cell problem. The field in the unit cell of the
structure can be solved using, for instance, the method of moments.
However, in-house and commercial softwares using this technique
consume considerable time and computational resources. This is
restrictive in the design of complex structures where the array
under study is only one of the building blocks. In these cases
accurate and simple analytical models thrive, for they are easy to
use and give good insight into physical phenomena.

For arrays of parallel capacitive strips located in a uniform host
medium, one can find in \cite{Marcuvitz} an equivalent circuit model
valid for oblique incidence of TE- and TM-polarized waves. Patch
arrays were not studied in \cite{Marcuvitz}. In \cite{Whites} it was
shown that the accuracy of the model suggested in \cite{Marcuvitz}
is lost when the gap between the strips is not very small compared
to the strip width. In \cite{DeLyser} the authors derived averaged
boundary conditions for planar arrays (in the first-order
approximation, where  the small parameter is the ratio
period/wavelength). However, the oblique incidence case was not
considered. In \cite{Compton_strips} a circuit model involving the
so-called approximate Babinet principle (the conventional Babinet
principle implies the location of two mutually complementing planar
grids in a uniform background material) was developed for strips
located at a dielectric interface. However, the analysis was
restricted to the normal incidence as in \cite{Whitbourn} and
\cite{Ulrich}. In \cite{Whitbourn} and \cite{Ulrich} periodical
arrays of conducting wires on a dielectric interface were studied.
In \cite{Holloway} the authors considered the reflection properties
of arrays of separate scatterers (an array of square patches was
studied as an example). The analysis was suitable for oblique
incidence (see also in \cite{Kuester}); however, the model
\cite{Holloway} is based on the assumption that the interaction of
scatterers is the dipole interaction. Therefore, the accuracy of
this model should be additionally studied for the case when the gaps
separating the patches or strips are narrow, and the capacitive
coupling between adjacent metal elements is strong. This situation
holds for many frequency selective surfaces and for all
high-impedance surfaces. When the effective capacitance of a planar
array is high, the resonance frequency of a high-impedance surface
comprising this array can be quite low. Structure miniaturization is
practically equivalent to the decreasing of the resonance frequency
of a structure with a given unit cell size. This target is always in
mind of designers of microwave structures. In the present paper we
will, in particular, consider the arrays with very small gaps
between metal elements.

In \cite{clavijo} the expressions for the input impedances of
high-impedance surfaces have been derived by treating the capacitive
grid layer and the metal-backed dielectric slab with embedded vias
as homogeneous materials with an anisotropic magneto-dielectric
tensor. The resulting expressions for the input impedances are
lengthy and complicated compared to the expressions of the present
paper. While deriving semi-intuitively the expressions for the
permittivities and permeabilities of the grid layer, the authors of
\cite{clavijo} touch the topic of the grid impedance of arrays of
patches without dealing with it in detail. In the present paper the
grid impedance is derived strictly from the known expressions of the
grid impedances for mesh of wires.

We start from the equivalent circuit model for arrays of parallel
capacitive strips and square patches located on a dielectric
interface. This model in the case of oblique incidence of plane
waves is based on the so-called averaged boundary conditions (ABC).
These conditions are derived below from the known averaged boundary
conditions formulated earlier for parallel inductive strips and
square strip meshes using the approximate Babinet principle (see
e.g. \cite{Tretyakov}). The obtained results are compared to the
ones found in the literature and to full-wave simulations. Then we
will do the same for impenetrable impedance surfaces. This approach
was already applied for the normal incidence in \cite{Tretyakov_1},
and for the oblique incidence in \cite{Simovski_1,Simovski}. In
\cite{Tretyakov_1} the authors studied the so-called mushroom
structures \cite{Sievenpiper}. In \cite{Simovski_1,Simovski}
high-impedance surfaces were based on self-resonant grids (planar
arrays of complex-shape elements). The accuracy for the oblique
incidence offered by the analytical model for arrays of
complex-shaped elements in \cite{Simovski_1,Simovski} was not high.
Analytical calculations served in \cite{Simovski_1,Simovski} for
heuristic purposes, and the Ansoft HFSS package was used for
quantitative calculations. In the present paper we consider arrays
of strips and patches and demonstrate the excellent accuracy of the
simple approach based on the equivalent circuit, averaged boundary
conditions, and the approximate Babinet principle. The reader should
notice, that this approach was outlined in book \cite{Tretyakov} for
arrays of metal patches. However, the results of \cite{Tretyakov}
for grids of patches are not accurate for oblique incidence, because
the periodicity in one of the tangential directions was not properly
taken into account.

\section{Grid impedance for capacitive strips and square patches}

\subsection{Array of patches}

\begin{figure}[t!]
\centering \subfigure[]{\epsfig{file=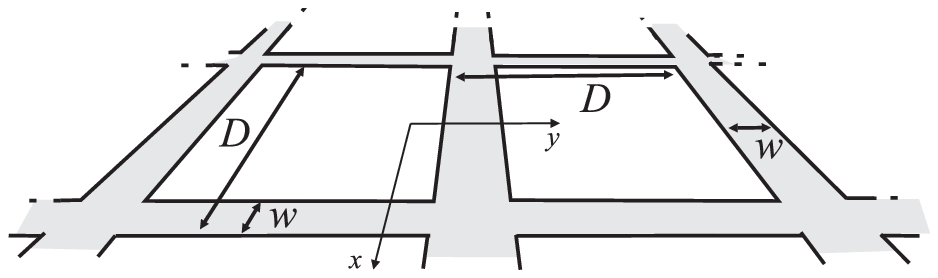, width = 7.5cm}}
\subfigure[]{ \epsfig{file = 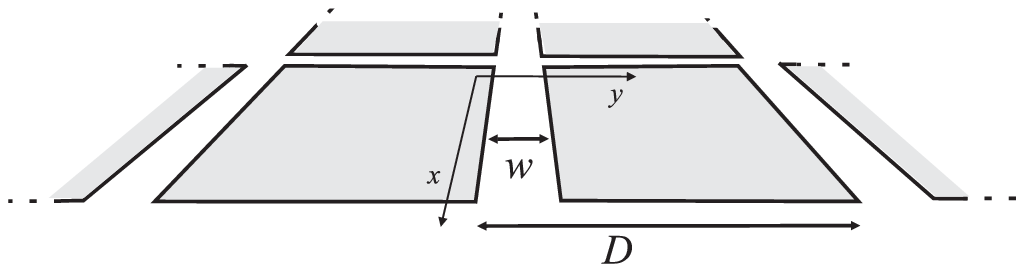, width = 7.5cm}}
 \caption{(a) A mesh of ideally conducting strips in homogeneous host
medium.
 (b) Array of patches in homogeneous host
medium. Metal parts are colored grey.} \label{fig:patches}
\end{figure}

First, we consider a mesh formed by laying parallel metallic strips
along the $x$- and $y$-axes, as shown in Fig.~\ref{fig:patches}(a).
If the period $D$ is electrically small, this structure is nearly
isotropic and its electromagnetic response only weakly depends on
the choice of the plane of incidence. As an example, assume
incidence in the $(x-z)$-plane. When $w\ll D$, where $w$ is the
width of the strips (see Fig.~\ref{fig:patches}(a)), this grid can
be considered as a mesh of strip wires with ideal contacts between
crossing wires. When the electric field has a non-zero $x$- or
$y$-component (parallel to the strips), the response of the grid is
inductive at low frequencies. This response can be characterized by
the so-called grid impedance, $Z_{\rm g}$, which relates the
averaged tangential component of the total electric field in the
grid $(x-y)$ plane $\widehat{E}_{\rm x}^{\rm tot}$ or
$\widehat{E}_{\rm y}^{\rm tot}$ to the averaged surface current
density $\widehat{J}$ induced on it by the incident plane wave and
flowing along the strips. For the TE-polarized case:
$\widehat{E}_{\rm y}^{\rm tot}=Z_{\rm g}^{\rm TE}\widehat{J}_{\rm
y}$, where $Z_{\rm g}^{\rm TE}$ is the grid impedance for the
TE-polarized incidence wave and $\widehat{J}_{\rm y}$ is the
averaged surface current density along the $y$-axis. The averaging
is made over the grid period $D$, and the current density
$\widehat{J}$ is equal to the jump of the $y$- or $x$-component of
the averaged magnetic field across the grid plane, respectively. The
averaged boundary conditions for such meshes (derived by M.I.
Kontorovich in the 1950s) can be found e.g. in \cite{Tretyakov}: \e
\widehat{E}_{\rm x}^{\rm tot} = j\frac{\eta_{\rm
eff}}{2}\alpha\left[\widehat{J}_{\rm x} + \frac{1}{k_{\rm eff}^2(1 +
\frac{b}{D})}\frac{b}{D}\frac{\partial^2}{\partial
x^2}\widehat{J}_{\rm x}\right], \label{eq:1} \f \e \widehat{E}_{\rm
y}^{\rm tot} = j\frac{\eta_{\rm eff}}{2}\alpha\widehat{J}_{\rm
y},\label{eq:2}\f for TM- and TE-polarized incident fields,
respectively. In the above formulae $b$ and $D$ are the periods of
the structure along the $x$- and $y$-axes, respectively, $\eta_{\rm
eff}=\sqrt{{\mu_0}/{\varepsilon_0\varepsilon_{\rm eff}}}$ is the
wave impedance of the uniform host medium with relative effective
permittivity $\varepsilon_{\rm eff}$ in which the grid is located,
and $\alpha$ is called the {\it grid parameter} \cite{Tretyakov}. In
our isotropic case $D=b$. Furthermore, $k_{\rm
eff}=k_0\sqrt{\varepsilon_{\rm eff}}$ is the wave number of the
incident wave vector in the effective host medium. $\mu_0$,
$\varepsilon_0$, and $k_0$ are the permeability, permittivity, and
the wave number in free space, respectively.

Though relations \eqref{eq:1} and \eqref{eq:2} were obtained for the
case when the host medium is uniform, we heuristically extend it to
the case when the grid is located on a surface of a dielectric
substrate with  the relative permittivity $\varepsilon_{\rm r}$. The
effective permittivity of an equivalent uniform medium in
\eqref{eq:1} and \eqref{eq:2} in this case can be approximated as
\cite{Compton_strips}: \e \varepsilon_{\rm eff} =
\frac{\varepsilon_{\rm r} + 1}{2}.\f

The grid parameter for an electrically dense ($k_{\rm eff}D\ll
2\pi$) array of ideally conducting strips reads \cite{Tretyakov}: \e
\alpha = \frac{k_{\rm eff}D}{\pi}\ln\left( \frac{1}{\sin{\frac{\pi
w}{2 D}}}\right),\l{alpha} \f where $w$  is the strip width. For
cases when $w\ll D$, the natural logarithm in relation \r{alpha} can
be approximated as $\ln\left(\frac{2D}{\pi w}\right)$. A more
accurate approximation for the grid parameter taking into account
terms of the order $(\frac{k_0D}{\pi})^5$ and oblique incidences can
be found in \cite{Yatsenko}. It has been shown in that paper that
for sparse wire grids the accuracy of the higher order approximation
is very good up to the frequencies when $D = \lambda$. Over this
limit the phase variations on the surface are too rapid and
averaging fails. At higher frequencies grating lobes emerge (i.e.
higher Floquet space harmonics become propagating) and anyway a
single impedance would not suffice for the description of the
properties of the surface.

Replacing $\frac{\partial}{\partial x}$ by $-jk_{\rm x}$, where
$k_{\rm x} = k_0\sin\theta$ ($\theta$ is the angle of incidence and
$k_{\rm y} = 0$) is the $x$-component of the incident wave vector in
the free space, we obtain from \eqref{eq:1} and \eqref{eq:2} the
grid impedances in the form: \e Z_{\rm g}^{\rm TM} =
j\frac{\eta_{\rm eff}}{2}\alpha\left(1 - \frac{k_0^2}{k_{\rm
eff}^2}\frac{\sin^2\theta}{2} \right),\label{eq:5} \f \e Z_{\rm
g}^{\rm TE} = j\frac{\eta_{\rm eff}}{2}\alpha.\label{eq:6} \f

Here, ${\rm TM }$ and ${\rm TE }$ refer to the TM- and TE-polarized
incident fields, respectively. Now we can derive the grid impedance
for the complementary structure, that is, for an array of patches
(see Fig.~\ref{fig:patches}(b)). To find its grid impedance we use
the approximate Babinet principle (e.g. \cite{Whitbourn, Tretyakov})
which reads in terms of grid impedances as: \e Z_{\rm g}^{\rm
TE}Z_{\rm g'}^{\rm TM} = \frac{\eta_{\rm eff}^2}{4},\label{eq:Bab}
\f where $Z_{\rm g'}^{\rm TM}$ is the grid impedance of the
complementary structure for TM-polarized incidence fields.
Similarly, the grid impedance of the complementary structure for
TE-polarized incident fields is obtained through the Babinet
principle and $Z_{\rm g}^{\rm TM}$. Applying the approximate Babinet
principle we have from \eqref{eq:5} and \eqref{eq:6}: \e Z_{\rm
g'}^{\rm TM} = -j\frac{\eta_{\rm eff}}{2\alpha},
\label{eq:TM_patches}\f \e Z_{\rm g'}^{\rm TE} = -j\frac{\eta_{\rm
eff}}{2\alpha\left(1 - \frac{k_0^2}{k_{\rm
eff}^2}\frac{\sin^2\theta}{2} \right)}. \l{TE_patches}\f

If the plane of incidence for the mesh of strips in
Fig.~\ref{fig:patches}(a) for the TE-polarized incident wave is the
$(x-z)$ plane, then (following from the Babinet principle) for the
array of patches in Fig.~\ref{fig:patches}(b) for the TE-polarized
incident field the plane of incidence is the $(y-z)$ plane.

\subsection{Arrays of strips}

\begin{figure}[t!]
\centering \subfigure[]{\epsfig{file = 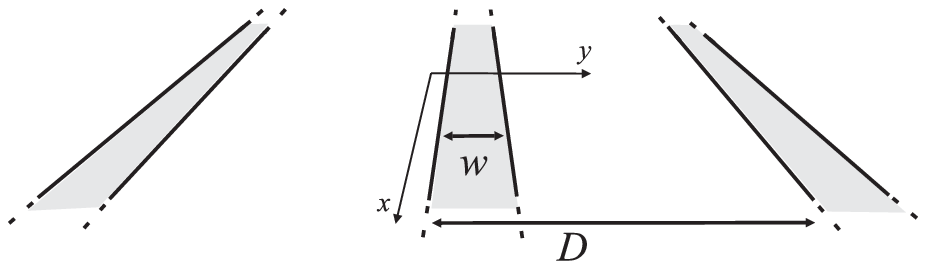, width = 7.5cm}}
\subfigure[]{ \epsfig{file=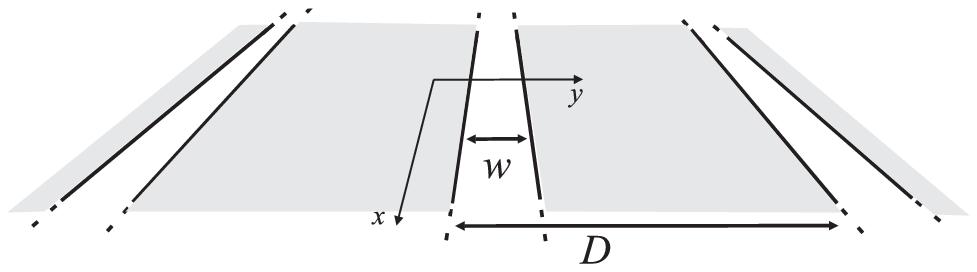, width = 7.5cm}} \caption{(a)
An inductive grid of metal strips in homogeneous host medium. (b) A
capacitive grid of metal strips in homogeneous host medium. Metal
parts are colored grey.} \label{fig:strips}
\end{figure}

Periodical grid of perfectly conducting strips whose width $w$ is
much smaller than the distance $D-w$ between the adjacent strips (we
again restrict the study by this case) is shown in
Fig.~\ref{fig:strips}(a) (current flows along the strips). The
expression for the grid impedance for both TE- and TM-polarizations
for this case can be found in \cite{Tretyakov}. The complementary
structure of Fig.~\ref{fig:strips}(a) is shown in
Fig.~\ref{fig:strips}(b) (current flows across the strips). Let the
electric field of the incident wave be $x$-polarized in
Fig.~\ref{fig:strips}(a). Then for the TM-polarized wave the
incidence plane is the $(x-z)$ plane ($k_{\rm y} = 0 $) and for the
TE-polarized wave the incidence plane is  the $(y-z)$ plane ($k_{\rm
x} = 0$). For the grid depicted in Fig.~\ref{fig:strips}(b) the
electric field is parallel to the $y$ axis. The TE polarization
corresponds to the incidence plane $(x-z)$ and the TM polarization
corresponds to the incidence plane $(y-z)$.

From \eqref{eq:Bab} and from the expressions for the grid impedances
for the structure depicted in Fig.~\ref{fig:strips}(a) (e.g. from
\cite{Tretyakov}) we get the grid impedance of the capacitive grid
of strips for both polarizations: \e Z_{\rm g'}^{\rm TM} =
-j\frac{\eta_{\rm eff}}{2\alpha},\label{eq:13} \f \e Z_{\rm g'}^{\rm
TE} = -j\frac{\eta_{\rm eff}}{2\alpha}\frac{1}{\left( 1 -
\frac{k_0^2}{k_{\rm eff}^2}\sin^2\theta\right)}.\label{eq:14} \f

These results correspond to those available in \cite{Marcuvitz}. The
structures presented in Figs.~\ref{fig:patches}(b) and
\ref{fig:strips}(b) have equivalent grid impedance for the
TM-polarized incident wave. For the TE polarization the grid
impedances of the capacitive grid of parallel strips differs from
that of the patch array due to the factor
$\frac{1}{2}\sin^2(\theta)$ in \eqref{eq:TE_patches} instead of
$\sin^2(\theta)$ in \eqref{eq:14} in the denominator.

\subsection{The reflection and transmission coefficients of patch arrays}

\begin{figure}[b!]
\centering \subfigure[]{\includegraphics[width = 7.5cm]{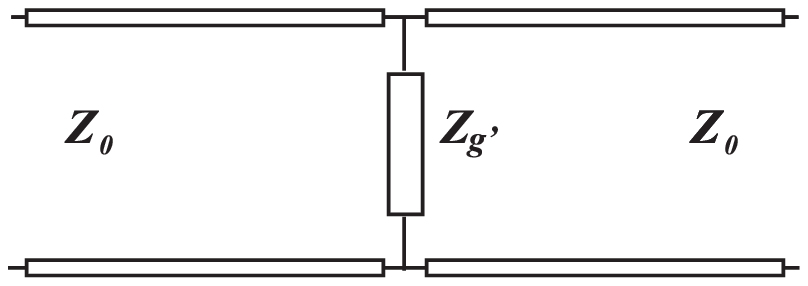}}
\subfigure[]{\includegraphics[width = 7.5cm]{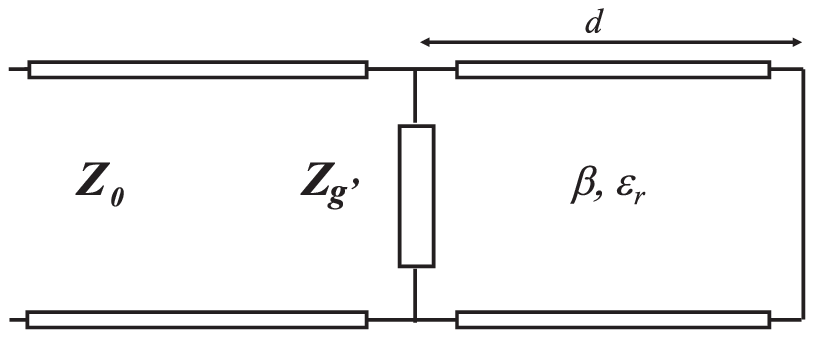}}
\caption{(a) The transmission-line model for a capacitive grid of
strips or an array of patches in free space. (b) The
transmission-line model for a high-impedance surface comprising a
grid of capacitive strips or an array of patches on top of a
metal-backed dielectric slab. The expressions for $Z_0$ and $Z_{\rm
g'}$ are given in \eqref{eq:Z0TE}, \eqref{eq:Z0TM}, and
\eqref{eq:Za_patches_fs}. \label{linja}}
\end{figure}

The equivalent circuit that can be used to calculate the reflection
and transmission coefficient from an array of patches with a certain
corresponding grid impedance $Z_{\rm g'}$ is shown in
Fig.~\ref{linja}(a) (see also e.g. in \cite{Marcuvitz}). The
transmission-line model for a high-impedance surface in
Fig.~\ref{linja}(b) will be discussed in the following section. The
free space impedances, $Z_0$, for the TE and TM modes are given for
different angles of incidence as: \e Z_0^{\rm TE} =
\frac{\eta_0}{\cos\theta}, \label{eq:Z0TE}\f \e Z_0^{\rm TM} =
\eta_0\cos\theta, \label{eq:Z0TM}\f where $\eta_0$ is the plane wave
impedance in free space. The input (surface) impedance referring to
the illuminated surface of the grid $Z_{\rm inp}^{\rm TE, TM}$ is a
parallel connection of the grid impedance $Z_{\rm g'}$ and $Z_0$: \e
Z_{\rm inp}^{-1}=Z_{\rm g'}^{-1}+Z_{0}^{-1}. \f

From \eqref{eq:TM_patches} and \eqref{eq:TE_patches} we have for the
case of an array of patches in free space: \e Z_{\rm g'}^{\rm TM} =
-j\frac{\eta_{\rm 0}}{2\alpha},\quad Z_{\rm g'}^{\rm TE} =
-j\frac{\eta_{\rm 0}}{2\alpha\left(1 - \frac{\sin^2\theta}{2}
\right)}. \label{eq:Za_patches_fs}\f The angular dependency of the
grid impedance in the TE-polarized case for an array of patches
located in free space in \eqref{eq:Za_patches_fs} is different from
the predictions of the known analytical model developed in
\cite{Holloway}. In \cite{Holloway} the reflection coefficients for
the TE- and TM-polarized cases are presented. The grid impedances
corresponding to the reflection coefficient in \cite{Holloway} would
read: \e Z_{\rm H}^{\rm TE,TM} = -j\frac{\eta}{2a},
\label{eq:Holloway_grid} \f
 where
\e a =
kD\frac{0.51\left(\frac{D-w}{D}\right)^3}{1-0.367\left(\frac{D-w}{D}\right)^3}.
\label{eq:Holloway_ref_TM1} \f The suffix ${\rm H}$ refers here to
Holloway et al. \cite{Holloway}. Apart from different approximations
for the grid parameter, the analytical results presented by
Eqs.~\eqref{eq:Za_patches_fs} and \eqref{eq:Holloway_grid} have
clearly different angular dependencies. In order to compare the
accuracy of the two models, in the following subsection we evaluate
the predictions of our analytical models and those in
\cite{Holloway} against full-wave simulation results.

\subsection{Numerical validation}

\begin{figure}[b!]
\centering \includegraphics[width = 7.5cm]{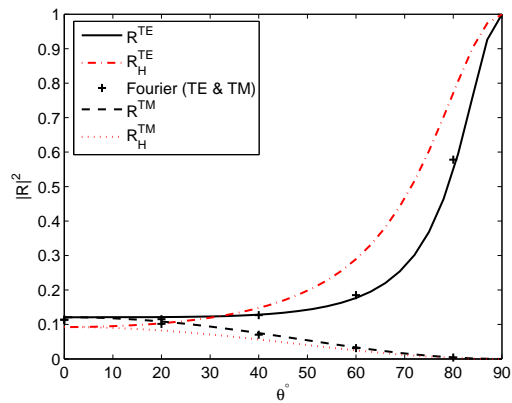} \caption{Color
online. The TE- and TM-reflection coefficient versus the incident
angle for an array of square patches. The dimensions of the array
are the following: $D=\lambda/10$ and $w = D/10$. $R_{\rm H}$
corresponds to the results according to the model by C. L. Holloway
et al. \cite{Holloway}. \label{results2}}
\end{figure}

Here we compare the analytical results obtained using the present
model and the model in \cite{Holloway} with simulations. The
numerical simulations are based on the Floquet expansion (the
Fourier modal method developed in \cite{Granet} and used in
\cite{Granet_stripgratings} for analysis of strip gratings). In
Fig.~\ref{results2} the results obtained using the Fourier modal
method are denoted as Fourier results. In all of the numerical
simulations the periodical structures were considered to be
infinite.

The reflectance of the patch array for the TE- and TM-polarized
incident fields is shown in Fig.~\ref{results2}. The reflection
coefficients have been calculated using the transmission-line model
in Fig.~\ref{linja}(a) and \eqref{eq:Za_patches_fs}. In
Fig.~\ref{results2} the results correspond to the dimensions
$D=\lambda/10$ and $w = D/10$. The analytical results are compared
with the results according to \eqref{eq:Holloway_grid}. These
results are denoted by $R_{\rm H}$ in Fig.~\ref{results2}. The
present analytical model shows very good agreement with the
numerical results, whereas the alternative analytical model
\cite{Holloway} appears to be significantly less accurate.

\section{High-impedance structures}

\subsection{Input impedance of high-impedance surfaces}

\begin{figure}[b!]
\centering \includegraphics[width = 7.5cm]{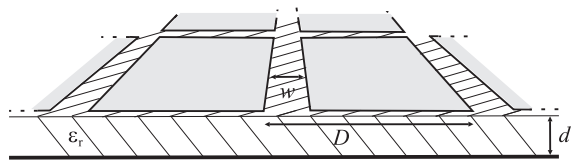} \caption{A
high-impedance structure consisting of an array of patches on top of
a metal-backed dielectric slab. \label{fig:mushroom}}
\end{figure}

The high-impedance surfaces (HIS) studied here consist of the above
considered capacitive grids or arrays on a metal-backed dielectric
slab with thickness $d$ and relative permittivity $\varepsilon_{\rm
r}$. In Fig.~\ref{fig:mushroom} a HIS with a patch array is shown.
Similarly, one can realize a HIS with parallel capacitive strips.
The input (surface) impedance of the HIS can be understood if we
look at Fig.~\ref{linja}(b). In Fig.~\ref{linja}(b) $\beta$ is the
propagation constant component orthogonal to the surface. The input
(surface) impedance is a parallel connection of the grid impedance
$Z_{\rm g'}$ and the surface impedance of the grounded dielectric
layer $Z_{\rm s}$: \e Z_{\rm inp}^{-1}={Z_{\rm g'}}^{-1}+Z_{\rm
s}^{-1}. \f The surface impedance for the oblique incidence can be
written in the dyadic form (see e.g. in \cite{Tretyakov}) which
splits into two scalars $Z_{\rm s}^{\rm TE}$ and $Z_{\rm s}^{\rm
TM}$ in the TE- and TM-cases, respectively: \e
\overline{\overline{Z}}_{\rm s} = j\omega\mu\frac{\tan\left(\beta
d\right)}{\beta}\left(\overline{\overline{I}}_{\rm t} -
\frac{\mathbf{k_{\rm t}}\mathbf{k_{\rm t}}}{k^2} \right),
\l{Zs_novias} \f where $\mu$ is the absolute permeability of the
substrate (in our case $\mu=\mu_0$), $\beta=\sqrt{k^2 - k^2_{\rm
t}}$, $k=k_0\sqrt{\varepsilon_{\rm r}}$ is the wave number in the
substrate material, and $\mathbf{k}_{\rm t}$ is the tangential wave
number component, as imposed by the incident wave.

For a HIS comprising capacitive strips the TM- and TE-input
impedances follow from Eqs.~\eqref{eq:13}, \eqref{eq:14} and
\r{Zs_novias}, and read, respectively: \e Z_{\rm c, inp}^{\rm TM} =
\frac{j\omega\mu\frac{\tan(\beta d)}{\beta}\cos^2(\theta_2)}{1 -
2k_{\rm eff}\alpha\frac{\tan(\beta d)}{\beta}\cos^2(\theta_2)},
\label{eq:Z_c_inp^TM}\f \e Z_{\rm c,inp}^{\rm TE} =
\frac{j\omega\mu\frac{\tan(\beta d)}{\beta}}{1 - 2k_{\rm
eff}\alpha\frac{\tan(\beta d)}{\beta}\left(1 -
\frac{2}{\varepsilon_{\rm r}+1}\sin^2\theta\right)},
\label{eq:Z_c_inp^TE}\f where $\theta$ is the angle of incidence and
$\theta_2$ is calculated from the law of refraction as: \e \theta_2
= \arcsin\left(\frac{\sin(\theta)}{\sqrt{\varepsilon_{\rm r}}}
\right).\f

In the case of patch arrays, the input impedances follow from
Eqs.~\r{TM_patches}, \r{TE_patches} and \r{Zs_novias}. For the
TM-polarization the input impedance equals to that of
\eqref{eq:Z_c_inp^TM}. For the TE-polarization the input impedance
reads: \e Z_{\rm p, inp}^{\rm TE} = \frac{j\omega\mu\frac{\tan(\beta
h)}{\beta}}{1 - 2k_{\rm eff}\alpha \frac{\tan(\beta
h)}{\beta}\left(1 - \frac{1}{\varepsilon_{\rm r} + 1}\sin^2\theta
\right)}. \label{eq:Z_p_inp^TE}\f

In the case of electrically thin substrates the expressions for the
input impedances can be further simplified by replacing the
$\tan(\beta h)$ by its argument $\beta h$. In addition, the
substrate losses can be taken into account in the analytical model
simply by replacing the relative permittivity with an appropriate
complex number. However, the effect of losses have been considered
to be out of the scope of this paper. Therefore in the following
part the validation of the model for HIS is conducted for lossless
structures.

\subsection{Numerical validation}

Since the HIS is impenetrable, in the absence of losses its
reflectance is equal to unity for the frequency range considered
here. Therefore the analytical model is verified from the reflection
phase diagrams for different angles of incidence. Our analytical
expressions are compared to the results of two accurate full-wave
simulations: Ansofts' High Frequency Structure Simulator 10.1.1.
(HFSS) and the Fourier modal method. In addition we have compared
our analytical expressions and the results of MoM simulations in the
case of square patches.

\begin{figure}[t!]
\centering \includegraphics[width = 7.5cm]{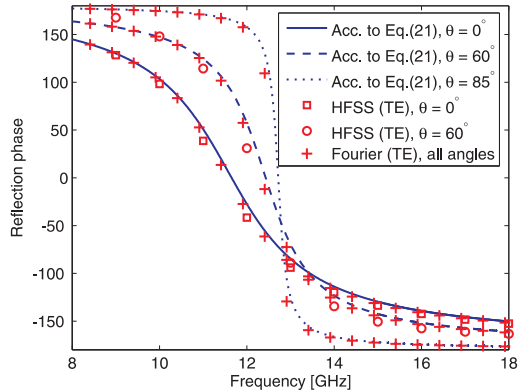} \caption{Color
online. The reflection phase diagram for TE-polarized incident
fields for a high-impedance surface consisting of capacitive strips
for different angles of incidence. The parameters of the
high-impedance surface are the following: $D = 2$\,mm, $w=0.2$\,mm,
$d=1$\,mm, and $\varepsilon_{\rm r} = 10.2$. HFSS simulations were
not done for $\theta=85^{\circ}$. \label{fig:results_TE_strips}}
\end{figure}

\begin{figure}[t!]
\centering \includegraphics[width = 7.5cm]{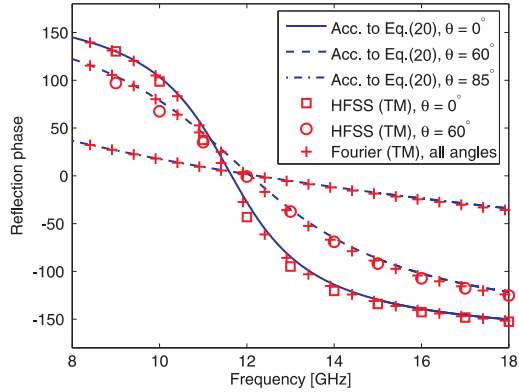} \caption{Color
online. The reflection phase diagram for TM-polarized incident field
for a high-impedance surface consisting of capacitive strips for
different angles of incidence. The parameters of the high-impedance
surface are the following: $D = 2$\,mm, $w=0.2$\,mm, $d=1$\,mm, and
$\varepsilon_{\rm r} = 10.2$. HFSS simulations were not done for
$\theta=85^{\circ}$. \label{fig:results_TM_strips}}
\end{figure}

In Figs.~\ref{fig:results_TE_strips} and \ref{fig:results_TM_strips}
the reflection phase diagram is presented for the TE- and
TM-polarized incident fields, respectively. It corresponds to a HIS
comprising capacitive strips on top of a dielectric slab grounded by
a perfectly conducting plane. The parameters of the HIS are the
following: $D = 2$\,mm, $w=0.2$\,mm, $d=1$\,mm, and
$\varepsilon_{\rm r} = 10.2$.

\begin{figure}[t!]
\centering \includegraphics[width = 7.5cm]{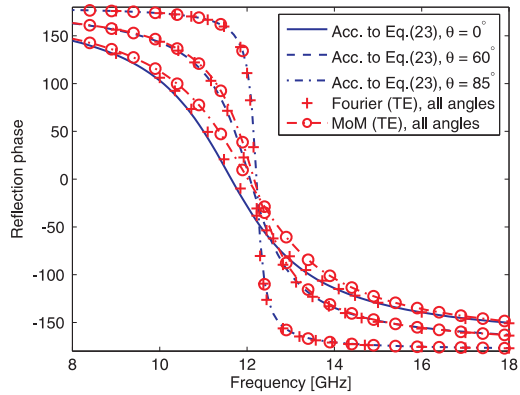} \caption{Color
online. The reflection phase diagram for TE-polarized incident field
for a high-impedance surface consisting of square patches for
different angles of incidence. The dimensions of the high-impedance
surface are the following: $D = 2$\,mm, $w=0.2$\,mm, $d=1$\,mm, and
$\varepsilon_{\rm r} = 10.2$. \label{fig:results_TE_patches}}
\end{figure}

\begin{figure}[t!]
\centering \includegraphics[width = 7.5cm]{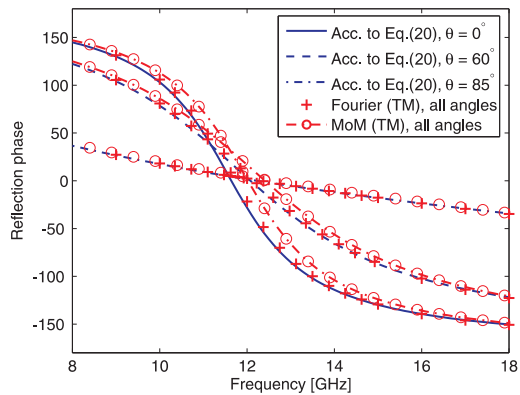} \caption{Color
online. The reflection phase for TM-polarized incident field for a
high-impedance surface consisting of square patches for different
angles of incidence. The dimensions of the high-impedance surface
are the following: $D = 2$\,mm, $w=0.2$\,mm, $d=1$\,mm, and
$\varepsilon_{\rm r} = 10.2$. \label{fig:results_TM_patches}}
\end{figure}

For the HIS comprising an array of square patches analogous results
are shown in Figs.~\ref{fig:results_TE_patches} and
\ref{fig:results_TM_patches}. The numerical parameters of this HIS
are the same as those in the previous example. The agreement between
the analytical and numerical results is very good for all angles of
incidence. The HIS comprising an array of patches was simulated also
with HFSS for the incidence angles of $0^\circ$ and $60^\circ$. The
results of the HFSS simulations showed excellent agreement with the
analytical and Fourier results.
Figures~\ref{fig:results_TE_strips}--\ref{fig:results_TM_patches}
show that the present analytical model is perfectly adequate for
lossless HIS comprising capacitive patch and strip arrays.

Let us define the resonant bandwidth of the HIS in the reflection
phase diagrams as the frequency band in which the reflection phase
is from $-90^\circ$ to $90^\circ$. It is clear that in
Figs.~\ref{fig:results_TE_strips}--\ref{fig:results_TM_patches} the
resonant bandwidth of the TM (TE) polarized wave is increased
(decreased) as the angle of incidence increases. Although the
surface impedance of the HIS changes with respect to the incident
angle, it does not explain this behavior. This is rather due to the
fact that the free space surface impedances in \eqref{eq:Z0TE} and
\eqref{eq:Z0TM} and their proportion to the surface impedance of the
HIS increase or decrease as the incident angle is increased,
respectively, when calculating the reflection coefficient.


For frequencies higher than $20$ GHz the HIS with the chosen
parameters cannot be homogenized and we cannot expect our model to
be adequate in this range. Using a more accurate approximation for
the grid parameter \cite{Yatsenko} the validity range of our model
could be extended. However, we cannot predict higher-order surface
impedance resonances with our model and the limits of the model have
not been studied here. Nevertheless, within the frequency band of
the main (lowest) resonance the present model has been shown to be
very accurate for all angles of incidence up to nearly grazing
incidence.

\section{Surface waves}

Knowledge about surface waves along the impedance surfaces is needed
when using the impedance surfaces as leaky wave antennas
\cite{Sievenpiper_surfacewaves}, artificial magnetic conductors for
antenna applications \cite{Goussetis}, or to suppress back radiation
of the antennas \cite{Maci}. In this section the propagation
properties of surface waves along a HIS are studied.


The derived expressions for the input impedance of the HIS can be
used for analysis of the surface wave propagation. Here we only
study the propagation of surface waves along the HIS surface
comprising an array of patches. The propagation properties of the
surface waves can be calculated from the transverse resonance
condition (e.g. \cite{Rozzi}): \e Z_0^{-1} + Z_{\rm p,inp}^{-1} = 0,
\label{eq:resonancecondition}\f where $Z_0$ is the free-space
impedance given in \eqref{eq:Z0TE} and \eqref{eq:Z0TM} for TE- and
TM-polarized incident fields, respectively, and the input impedance
of the surface $Z_{\rm p,inp}$ is given in \eqref{eq:Z_c_inp^TM} and
\eqref{eq:Z_p_inp^TE} for TM- and TE-polarized incident fields,
respectively. In the absence of losses $Z_{\rm p,inp}$ is purely
imaginary. In this case $Z_0$ needs to be purely imaginary in order
for \eqref{eq:resonancecondition} to be true. This means that the
angle of incidence in \eqref{eq:Z0TE} and \eqref{eq:Z0TM} needs to
be imaginary, that is, $\cos(\theta) = \frac{\sqrt{k_0^2 - k_{\rm
t}^2}}{k_0}$ and $k_{\rm t}
> k_0$ (waves along the $z$-axis in free space need to attenuate).

\begin{figure}[t!]
\centering \includegraphics[width = 7.5cm]{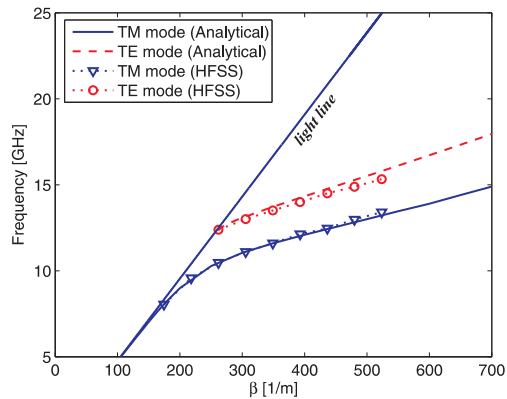} \caption{Color
online. The propagation properties of the surface waves along the
example HIS. The dimensions of the high-impedance surface are the
following: $D = 2$\,mm, $w=0.2$\,mm, $d=1$\,mm, and
$\varepsilon_{\rm r} = 10.2$. \label{fig:dispersiondiagram}}
\end{figure}

As an example, the dispersion of surface waves propagating along the
HIS studied in the previous section are shown in
Fig.~\ref{fig:dispersiondiagram}. The dimensions of the HIS are the
following: $D = 2$\,mm, $w=0.2$\,mm, $d=1$\,mm, and
$\varepsilon_{\rm r} = 10.2$. The simulations have been done with
HFSS 10.1.1 for an infinite structure. The analytical results agree
very well with the simulated results.

\section{Conclusions}

In this paper we have considered two types of planar capacitive
grids of metal elements separated by thin slits from one another,
namely grids of strips and arrays of square patches. The equivalent
grid impedance for the arrays of square patches has been derived. We
have shown that the grid impedance for strips and patches have
different angular dependencies. The obtained expressions based on
the transmission-line approach and the Babinet principle have been
thoroughly checked using both commercial simulation software and
in-house numerical codes based on the full-wave equations. Careful
comparison have demonstrated that the present model agrees with
results of \cite{clavijo} and is more accurate compared to the other
previously reported analytical models. We must emphasize here that
the Fourier Modal Method allows the increase of the spatial
resolution around the edges of the strips or patches, where a
singularity of the field occurs. As a consequence, the algorithm
does not suffer from instability. It follows from this that, while
checking the convergence of the results carefully, 21 Floquet
harmonics in one direction are enough to obtain good accuracy with
this method.

Furthermore, we have applied these analytical results for
high-impedance surfaces, where a planar grid is located over a
grounded dielectric layer. Here the comparison with the exact
solutions allowed us to check the assumption that the same grid
impedance as derived for the uniform host medium can be applied when
the grid is located at a dielectric interface (using the concept of
the effective permittivity). For two types of HIS under study the
verification against numerical results has demonstrated very good
agreement in the frequency range of the main surface resonance,
where the HIS operates as an artificial magnetic conductor. This
conclusion is true for both polarizations and all angles of
incidence up to nearly grazing incidence. It has also been shown in
this paper that the proposed model for high-impedance surfaces
allows very accurate prediction of the surface wave propagation for
both polarizations. The present study allows us to conclude that the
simple transmission-line model combined with the approximate Babinet
principle and the assumption of effective uniform permittivity (when
calculating the grid impedance of the capacitive array) is, in fact,
a very accurate model for these high-impedance surfaces.

\section*{Acknowledgements}

The work was supported in part by the Academy of Finland and Tekes
through the Center-of-Excellence Programme. Olli Luukkonen wishes to
thank the Nokia Foundation for financial support and acknowledge Dr.
Sergey Dudorov for useful discussions.

\end{document}